\documentclass[authoryear,12pt]{elsarticle}

\usepackage[english]{babel}
\usepackage{natbib}
 \bibpunct[, ]{(}{)}{,}{a}{}{,}%

\usepackage[nolist]{acronym}	
\usepackage[section]{placeins} 
\usepackage[tracking=true]{microtype}  
\usepackage{amsmath} 
\usepackage{mathtools} 
\usepackage{booktabs} 
\usepackage{siunitx} 
\usepackage{multirow} 
\usepackage{lscape} 
\usepackage{xcolor}

\usepackage{threeparttable}
\usepackage{enumitem}

\usepackage[linesnumbered,ruled,vlined,norelsize]{algorithm2e}
\SetKwRepeat{Do}{do}{while}

\SetCommentSty{mycommfont} 

\usepackage{afterpage} 

\newcolumntype{H}{>{\setbox0=\hbox\bgroup}c<{\egroup}@{}} 

\usepackage[hidelinks,breaklinks=true]{hyperref}

\usepackage{amsmath}
\usepackage{mathrsfs}
\usepackage{amssymb}

\newcommand{\etevrp}{\ac{E2E-VRP}}

\newcommand{\ft}{$1^\text{st}$}      
\newcommand{\sd}{$2^\text{nd}$}      

\newcommand{\setv}{N}           
\newcommand{\Ns}{N_S}           
\newcommand{\nums}{n_s}
\newcommand{\Nc}{N_C}           
\newcommand{\numc}{n_c}
\newcommand{\Nr}{N_R}           
\newcommand{\numr}{n_r}

\newcommand{\cH}{\bar{H}}
\newcommand{\cN}{\bar{N}}
\newcommand{\cA}{\bar{A}}

\newcommand{\numvf}{m^1}        
\newcommand{\capvf}{Q_1}        
\newcommand{\numvsk}{m_k}       
\newcommand{\caps}{B_k}         
\newcommand{\capvs}{Q_2}        
\newcommand{\fvcostA}{U_1}      
\newcommand{\fvcostB}{U_2}      

\newcommand{\ltot}{L}           

\newtheorem{prop}{Proposition}

\newcommand{\lc}[1]{\mathscr{#1}}

\newcommand{\mdcvrp}{\ac{MDC-VRP}}

\newcommand{\lbzero}{LB0}
\newcommand{\lbuno}{LB1}
\newcommand{\lbdue}{LB2}

\newcommand{\deltap}[1]{\Delta(#1)}

\begin{document}

\title{The Electric Two-echelon Vehicle Routing Problem}

\author[vie]{U.~Breunig}
\ead{ulrich.breunig@univie.ac.at}

\author[bol]{R.~Baldacci}
\ead{r.baldacci@unibo.it}

\author[vie]{R.F.~Hartl}
\ead{richard.hartl@univie.ac.at}

\author[rio]{T.~Vidal}
\ead{vidalt@inf.puc-rio.br}

\address[vie]{Department of Business Administration, University of Vienna, Austria}
\address[bol]{Department of Electrical, Electronic, and Information Engineering ``Guglielmo Marconi'', University of Bologna, Italy}
\address[rio]{Departemento de Inform\'{a}tica, Pontif\'{i}cia Universidade Cat\'{o}lica do Rio de Janeiro, Brazil}

\begin{abstract}
Two-echelon distribution systems are attractive from an economical standpoint and help to keep large vehicles out of densely populated city centers. Large trucks can be used to deliver goods to intermediate facilities in accessible locations, whereas smaller vehicles allow to reach the final customers. Due to their reduced size, pollution, and noise, multiple companies consider using an electric fleet of terrestrial or aerial vehicles for last-mile deliveries.

Route planning in multi-tier logistics leads to notoriously difficult problems. This difficulty is accrued in the presence of an electric fleet since each vehicle operates on a smaller range and may require planned visits to recharging stations. To study these challenges, we introduce the \emph{\ac{E2E-VRP}} as a prototypical problem. We propose a \ac{LNS} metaheuristic as well as an exact mathematical programming algorithm, which uses decomposition techniques to enumerate promising first-level solutions in conjunction with bounding functions and route enumeration for the second-level routes. These algorithms produce optimal or near-optimal solutions for the problem and allow us to evaluate the impact of several defining features of optimized battery-powered distribution networks.

We created representative \ac{E2E-VRP} benchmark instances to simulate realistic metropolitan areas. In particular, we observe that the detour miles due to recharging decrease proportionally to $1/\rho^x$ with $x \approx 5/4$ as a function of the charging stations density $\rho$; e.g., in a scenario where the density of charging stations is doubled, recharging detours are reduced by 58\%. Finally, we evaluate the trade-off between battery capacity and detour miles. This estimate is critical for strategic fleet-acquisition decisions, in a context where large batteries are generally more costly and less environment-friendly.
\end{abstract}


\maketitle

\section{Introduction}
\label{section:introduction}

Nowadays, as technology for electric mobility progresses, multi-tier delivery schemes are naturally destined to make use of electric vehicles, and multiple companies have, in practice, already operated this transition \citep{Foltynski2014}. Yet, electric vehicles also pose specific challenges, due to their limited autonomy, smaller capacity, and the possible need of planned visits to charging stations. Moreover, whereas early charging technologies required several hours for a full recharge, recent developments of fast-charging or battery-swap stations \citep{Yang2015,Hof2017,Keskin2018} allow energy replenishment in half an hour. The growing adoption of these technologies allows en-route recharging (e.g., during lunch breaks) in metropolitan distribution systems, as well as the use of cheaper lightweight vehicles \citep{Perboli2018a,Perboli2018b} with smaller batteries. Last but not the least, the study on battery-powered distribution is not limited to terrestrial vehicles, but also meets critical applications in last-mile distribution using aerial vehicles~(i.e.,~drones -- \citealt{Poikonen2017} and \citealt{Wang2017a}), which typically have a smaller autonomy.

To focus on these challenges, we introduce the \ac{E2E-VRP} as a prototypical problem. It is a natural extension of the \ac{2E-VRP} in which electric vehicles are used on the second echelon. Given a geographically-dispersed set of customers demanding an amount of a single commodity, a set of satellites (intermediate facilities), a set of charging stations, and a central depot where the commodity is kept, the \ac{E2E-VRP} seeks least-cost delivery routes to transport the commodity from the depot to the satellites with conventional vehicles (first-level), and from the satellites to the customers using an electric fleet (second-level).
Some additional rules must be satisfied with respect to the basic \ac{2E-VRP}:
\begin{itemize}[nosep,leftmargin=*]
\item Electric vehicles have a limited \textit{driving range}, which can be fully replenished at a charging station;
\item Each second-level route originates at a satellite, visits a sequence of customers and possibly some \textit{recharging stations}, and returns to the same satellite;
\item Each satellite also hosts a charging station at its location;
\item Charging stations can be used multiple times, but a consecutive visit to two charging stations in a second-level route is prohibited.
\end{itemize}
The cost of the solution, to be minimized, includes a fixed cost for each vehicle in use, as well as driving costs proportional to the distance traveled.

To solve this problem, we introduce a \ac{LNS}-based metaheuristic which combines a restricted set of destruction and reconstruction operators, a local search procedure, and a fast labeling algorithm to optimize the visits to charging locations. We also propose an exact mathematical programming algorithm, which uses a decomposition technique to enumerate promising first-level solutions along with bounding functions and route enumeration for the second level, using problem-tailored pricing algorithms. These two methods can be viewed as extensions of the approaches of \cite{Breunig2016} and \cite{Baldacci2013} for the classical \ac{2E-VRP}, in which specialized route evaluation techniques, labeling algorithms and dominance strategies have been integrated to efficiently manage the selection of recharging stations for the electric vehicles.

Not only do these algorithms allow to find optimal or near-optimal solutions for the \ac{E2E-VRP}, and thus respond to the need of advanced algorithms for future city-logistics planning, but they also open the way to an analysis of several defining features of optimized battery-powered city-distribution networks. To that end, we created new datasets which simulate the general characteristics of a metropolitan area, and examine the impact of the density of the charging station network and the capacity of the vehicles'  batteries on the cost-efficiency of the optimal solutions of the problem.

The remainder of this paper is organized as follows.
Section~\ref{section:review} reviews the related literature and Section~\ref{section:description} formally describes the problem.
Then, Sections~\ref{sec:exm} and \ref{section:method} describe, respectively, the proposed exact and heuristic algorithms. Section~\ref{section:results} reports our computational experiments and sensitivity analyses, and Section~\ref{sec:conc} concludes.

\section{Related Literature}
\label{section:review}

We review the existing solution algorithms for the \acp{2E-VRP}, discuss the recent studies dedicated to routing optimization for vehicles with alternative fuels, and finally examine the use of en-route recharging in recent studies and applications. \\

\noindent
\textbf{Two-echelon vehicle routing problems.}
Several early studies focused on mathematical programming solution techniques for the 2EVRP. \citet{feliu2007two} were the first to describe a branch-and-cut algorithm based on a commodity flow formulation that solved instances with up to $32$ customers and $2$ satellites.
The method of \citet{feliu2007two} was improved by \citet{perboli2010new} and \citet{Perboli2011} by adding valid inequalities in a cutting plane fashion.
Optimal solutions for instances with up to $32$ customers and $2$ satellites were found by the method of \citet{Perboli2011}.
\citet{Jepsen2012} described a branch-and-cut algorithm based on a new mathematical formulation and different valid inequalities.
Exact algorithms were also designed by \citet{Baldacci2013} and by \citet{Santos2015}.
\citeauthor{Santos2015} described a branch-and-cut-and-price algorithm for the 2EVRP that relies on a reformulation based on the $q$-routes relaxation proposed for the \ac{CVRP} by \citet{christofides_exact_1981}. \citeauthor{Baldacci2013} proposed an exact method for solving the 2EVRP based on a set partitioning formulation with side constraints. They described a bounding procedure that is used by the exact algorithm to decompose the  problem into a limited set of \acp{MDC-VRP} with side constraints. The optimal \ac{2E-VRP} solution is obtained by solving the set of \acp{MDC-VRP} generated. The method was tested on 207 instances, taken both from the literature and newly generated, with up to 100 customers and 6 satellites. The results obtained by \citet{Baldacci2013} show that their exact algorithm outperforms the existing methods from the papers described above. Finally, \cite{Perboli2018} recently found new valid inequalities for the \ac{2E-VRP}. Using these inequalities within a branch-and-cut algorithm allowed to solve several new instances with up to 50 customers.

The number of heuristics, metaheuristics and case studies focused on multi-echelon vehicle routing problems has also rapidly grown in the last decade. The surveys by \cite{Cuda2014a} and \cite{Schiffer2018} capture well the breadth of this line of research.  The former covers different two-echelon structured transportation problems: \acp{2E-LRP}, \acp{2E-VRP} and \acp{TTRP}. The latter focuses on vehicle routing problems and location routing problems with intermediate stops, dedicated to replenishment, refueling or idling. Among the most recent contributions in this domain,
\citet{Zeng2014a} proposed a greedy randomized adaptive search procedure with a route-first cluster-second splitting algorithm and a variable neighborhood descent for the \ac{2E-VRP}. Their results are promising, but the algorithm was only tested on the smaller benchmark instances with up to 50 delivery points.
\citet{Breunig2016} introduced a \ac{LNS} for \acp{2E-VRP} and the \ac{2E-LRPSD}. The method uses six destroy and one repair operator as well as some well-known local search procedures. It finds or improves 95\% of the best known solutions for the classical benchmark instances. Given the efficiency and the effectiveness of this method, the same general structure has been used for the heuristic proposed in this paper, in addition with multiple improvements and adaptations to account for the specificities of electric vehicles. Later on, \citet{Wang2017} studied an extension of the \ac{2E-VRP} with stochastic demands, described as a stochastic program with recourse. A genetic algorithm was proposed, and the results on the problem with stochastic demands were compared to the best known deterministic solutions.\\

\noindent
\textbf{Electric vehicle routing, en-route recharging and battery swaps.}
Over the last decade, research has also rapidly progressed on \acp{VRP} with alternative propulsion modes: electric or hybrid. As generally reflected in the surveys of \cite{Montoya2016,Pelletier2016} and \cite{Schiffer2018}, many of these studies consider possible en-route recharging or battery swaps to overcome the range limitations of electric vehicles.

\citet{conrad2011} were amongst the first to consider optimization techniques for electric vehicles and possible recharging stops. In the proposed \emph{recharging} \ac{VRP}, batteries can be charged at customer locations subject to additional costs. Other studies were focused on \acp{VRP} considering different aspects of environment-friendly transport. In particular, \cite{Erdogan2012} proposed the \ac{GVRP}, involving battery-powered vehicles with possible en-route recharging at dedicated stations, and evaluated the implications of refueling-stations availability and dispersion.

After these seminal works, the literature progressed towards more intricate problem variants and solution methods. \cite{Schneider2013} introduced additional time-window constraints for customer deliveries as well as recharging delays. New benchmark instances were introduced, and solved by means of a hybrid heuristic combining \ac{VNS} and \ac{TS}. \citet{Desaulniers2016} developed an exact method based on branch-price-and-cut, and presented computational results for the same benchmark instances.

Moreover, to progress towards real applications and improve the accuracy of the studies, other important characteristics of real delivery networks have been considered.
Heterogeneous fleets with different propulsion modes were studied in \cite{felipe2014}, jointly with a heterogeneous set of recharging stations with different cost and recharging time. \citet{goeke2015} considered a mix of conventional  and electric vehicles, evaluating the energy consumption of an electric vehicle as a function of speed, gradient and cargo load distribution. \citet{Hiermann2016} proposed the \ac{EFSMFTW}, in which the deliveries can be performed with a mix of vehicle types, differing in their acquisition cost, freight capacity, and battery size. Experiments were conducted with a branch-and-price algorithm and a hybrid heuristic. This research was extended in \cite{Hiermann2017} to study the impact of additional plug-in hybrid vehicles. \citet{keskin2016} introduced an \ac{ALNS} for a problem variant in which partial recharging is allowed.

In a recent case study, \cite{wang2017b} stressed several interesting facts regarding the use of large battery powered commercial vehicles. They show that in the real-world transit network based in Davis, California, range anxiety can be mitigated by adopting good recharging strategies, and that several companies adapt much stricter range limits than the theoretical ones (typically half of the range) to extend battery life cycles. Finally, they indicate a list of case studies in which battery swapping has been applied to electric bus transit systems, a strategy which is justified by the fact that fast charging techniques can significantly extend vehicle ranges within only five to ten minutes of charging. Other recent works have proposed to optimize fast or partial charging strategies. In particular, \cite{Schiffer2017} introduced the electric location routing problem with time windows and possible partial recharging stops, whereas \cite{Keskin2018} extended the \ac{EVRPTW} with different types of stations (super-fast, fast, and normal ones) for en-route recharging.\\

\noindent
\textbf{Real-world applications of en-route recharging.}
As early as 2008, electric buses were installed for the Beijing Olympic Games. Each bus, of a capacity of 50 seats, can undergo battery swapping up to three times a day \citep{evworld2008}.
Similarly, fully electric public transit buses have been operating in Vienna, Austria for several years now. These buses stop for approximately 15 minutes at the end of their route to traverse the inner city and recharge the batteries with an overhead system multiple times a day \citep{wrlinien2013}. \cite{Schiffer2016} also recently established a case study of  ``TEDi Logistik GmbH \& Co. KG'', which provides freight transportation services and relies on two electric vehicles. They concluded that the operations would be significantly improved if fast-charging stations were additionally available at a few customers locations for en-route recharging, therefore allowing to reach locations located further than 70km from the depot.
Finally, \url{JD.com}, one of China's biggest e-commerce companies, is replacing its existing fleet with fully electric vehicles \citep{gizmochina2017}. Seeing the need for it, they recently launched a contest for optimization algorithms capable of routing vehicles with en-route recharging \citep{jdata2018}.

Our study shares the same objectives as many aforementioned papers: bridging the gap between academic electric \acp{VRP} and real problem attributes. The current literature on electric vehicles has only considered simplistic delivery networks with a single depot and a single echelon, but it is well known that city logistics usually involve richer configurations, with interconnected echelons and transportation modes. Moreover, the restricted range of the electric vehicles and their possible need for en-route recharging bring new challenges which have to be considered when selecting intermediate facilities (i.e., satellite) locations. We therefore propose to study the impact of electrical fleets in second-level routes, in a two-echelon delivery setting, where electric vehicles are likely to be needed.

\section{Problem Description}
\label{section:description}

The \ac{E2E-VRP} addressed in this paper can be formally described as follows.

A mixed graph $G=(\setv,E,A)$ is given, where the vertex set $\setv$ is partitioned as $\setv=\{0\} \cup \Ns \cup \Nc \cup \Nr$. Vertex $0$ represents the depot, $\Ns=\{1,2,\dots,\nums\}$ represents $\nums$ satellites, $\Nc=\{\nums+1,\dots,\nums+\numc\}$ represents $\numc$ customers, and $\Nr=\{\nums+\numc+1,\dots,\nums+\numc+\numr\}$ represents $\numr$ charging stations. The edge set $E$ is defined as $E=\{\{0,j\} \, : \, j \in \Ns\} \cup \{\{i,j\}: i, j \in\Ns, i < j \}$ and the arc set $A$  as $A=\{(i,j) \, : \, i,j \in \Ns \cup \Nc \cup \Nr, i\neq j \} \setminus \{(i,j): i,j \in \Ns\} \setminus \{(i,j): i,j \in \Nr\}$.
A travel or routing cost $d_{ij}$ is associated with each edge $\{i,j\} \in E$ and with each arc $(i,j) \in A$.

Each customer $i \in \Nc$ requires a supply of $q_i$ units of goods to be delivered from the depot using the following two types of vehicles. A fleet of $\numvf$ vehicles of capacity $\capvf$ located at depot $0$ and a fleet of $\numvsk$ vehicles of capacity $\capvs < \capvf$ located at satellite $k \in \Ns$. Moreover, at most $m^2 \leq \sum_{k \in \Ns} \numvsk$ second-level vehicles can be used.

A \ft-level vehicle route is a simple cycle in $G$ passing through the depot and a subset of satellites such that the total demand delivered is less than or equal to $\capvf$. A satellite $k \in \Ns$ can be visited by more than one \ft-level route and has a capacity $\caps$ that limits the demand that can be delivered to it.

A \sd-level route is a circuit in $G$ passing through a satellite and a subset of customers and charging stations and such that the total demand of the visited customers does not exceed the vehicle capacity $\capvs$ and the following charging station constraints are respected. Each vehicle on the \sd-level has a maximum battery capacity $\ltot$, and a battery consumption $c_{ij}$ is associated with each arc $(i,j) \in A$; the maximum battery consumption of a vehicle without a visit to a charging station is therefore equal to $\ltot$. Charging stations can be visited right after or before a satellite, or in between customers, and, whenever a charging station is visited, a vehicle is fully charged up to level $\ltot$. In the scope of this short-haul problem, we prohibit a consecutive visit to two charging stations.

Fixed costs $\fvcostA$ and $\fvcostB$ are also associated with the use of \ft-level and \sd-level vehicles, respectively. The cost of a route (\ft-level or \sd-level) is equal to the sum of the costs of the traversed edges or arcs plus the fixed cost.

The problem asks to design both \ft-level and \sd-level routes so that the quantity delivered from each satellite is equal to the quantity received from the depot, each customer is visited exactly once, and the total cost of the routes is minimized.\\

\noindent
\textbf{Multigraph reformulation.}
The \ac{E2E-VRP} can be reformulated as a routing problem on a multigraph $G'=(N',E',A')$, where $N'=\{0\} \cup \Ns \cup \Nc$ is the vertex set, $E'=\{\{0,j\} \, : \, j \in \Ns\} \cup \{\{i,j\}: i, j \in\Ns, i < j \}$ is the edge set and $A'$ is the arc set. Arc set $A'$ is used to represent \sd-level routes and is defined as $A'= \{(i,j): i,j \in \Ns \cup \Nc, i\neq j\} \setminus \{(i,j):i, j \in \Ns \}$. The arc set $A'$ also contains the following set of arcs:
\begin{itemize}
  \item With each arc $(i,j) \in A'$ are associated $h(i,j)$ arcs representing the different paths that a \sd-level vehicle can take to go from vertex $i$ to vertex $j$ with at most one charging station visited in between vertices $i$ and $j$.
  \item A \textit{cost} $d(i,j,p)$, a \textit{consumption} $c(i,j,p)$ and a \textit{charging station} $s(i,j,p) \in \Nr$ are associated with each arc $(i,j,p)$, $p=1,\dots,h(i,j)$, $\forall (i,j) \in A'$. We assume that $s(i,j,p)=0$ if arc $(i,j,p)$ represents the direct path $(i,j)$ without any charging station visited in between $i$ and $j$.
  \item The cost $d(i,j,p)$ and the consumption $c(i,j,p)$ of arc $(i,j,p)$ are defined as follows:
\begin{equation*}
            \left \{
                \begin{array}{ll}
                  d(i,j,p)=d_{ij}, \; c(i,j,p)=c_{ij}, & \quad \text{ if } s(i,j,p)=0  \vspace*{0.2cm} \\
                  d(i,j,p)=d_{ik}+d_{kj}, \; c(i,j,p)=c_{kj}, & \quad \text{ if } k=s(i,j,p) \neq 0.
                \end{array}
            \right .
        \end{equation*}
\end{itemize}

Multigraph $G'$ does not contain any arc $(i,j,p)$ such that $s(i,j,p)=0$ and $c_{ij} > \ltot$, or  $k=s(i,j,p)\neq 0$ and $c_{ik} > \ltot$ or $c_{kj} > \ltot$. Notice that for arc $(i,j,p)$ with $s(i,j,p) \neq 0$ value $\ltot-c(i,j,p)$ represents the battery level of the vehicle after arriving  at vertex $j$ whereas if $s(i,j,p)=0$, i.e., no charging station is visited in between $i$ and $j$, the battery level at vertex $j$ is equal to $b-c(i,j,p)$, where $b$ is the battery level at vertex $i$.

A \sd-level route for satellite $k \in \Ns$ in graph $G'$  is a simple circuit in $G'$ passing through a satellite and a subset of customers and such that (i) the total demand of the visited customers does not exceed the vehicle capacity $\capvs$ and (ii) the vehicle leaves satellite $k$ with a consumption equal to 0 (or, equivalently, the vehicle is fully charged) and its consumption at each visited vertices does not exceed the maximum battery capacity $\ltot$.

The following proposition holds.
\begin{prop}
 There is a one-to-one correspondence between \sd-level routes in $G$ and \sd-level routes in $G'$.
\end{prop}

Moreover, the set of arcs $A'$ can be reduced by means of the following dominance rule.
\begin{prop}
 An optimal \etevrp\ solution cannot contain an arc $(i,j,r_1)$ if there exists another arc $(i,j,r_2)$, $r_1 \neq r_2$, such that:
 \begin{enumerate}[nosep]
   \item $i \in \Ns$: $d(i,j,r_1) \geq d(i,j,r_2)$ and $c(i,j,r_1) \geq c(i,j,r_2)$;
   \item $i \in \Nc$: $d(i,j,r_1) \geq d(i,j,r_2)$ and $c(i,j,r_1) \geq c(i,j,r_2)$, and $c_{ik_1} \geq c_{ik_2}$, $k_1=s(i,j,r_1)$, $k_1 \neq 0$, and $k_2=s(i,j,r_2)$, $k_2 \neq 0$.
 \end{enumerate}
\end{prop}

\section{Solving the \etevrp\ to Optimality}
\label{sec:exm}

The method used to solve the \ac{E2E-VRP} to optimality is based on the exact method proposed by \citet{Baldacci2013} for the \ac{2E-VRP}.
More precisely, we tailored the method described by \citeauthor{Baldacci2013} to handle the multigraph $G'$  described in Section~\ref{section:description}.
The exact method consists of the following two main steps.
\begin{enumerate}
\item The set of all \ft-level routes is generated and a lower bound \lbzero\ on the \etevrp\ is computed. The computation of \lbzero\ is based on a  integer relaxation that results in a multiple-choice knapsack problem. In computing lower bound \lbzero, we extended the $ng$-routes relaxation used in \citeauthor{Baldacci2013} to the case of our multigraph $G'$ (see below).
\item The set of all possible subsets of \ft-level routes that could be used in any optimal \etevrp\ solution is generated. For each subset of \ft-level routes the following steps are executed:
        \begin{enumerate}
            \item[(i)] Lower bound \lbzero\ is computed by fixing the selected set of \ft-level routes in solution.
            If the resulting lower bound is greater than or equal to the cost of the best incumbent \ac{E2E-VRP} solution, then the current subset is rejected, otherwise the next step is executed;
            \item[(ii)] The \ac{E2E-VRP} problem obtained by considering only the  selected set of \ft-level routes is solved to optimality.
            The resulting problem is a \mdcvrp, that is solved using the method proposed by \cite{Baldacci2009}. The optimal solution cost of the \etevrp\ corresponds to the minimum solution cost of such \mdcvrp s. In solving problem \mdcvrp, we extended the procedure used to generate elementary routes described in \cite{Baldacci2009} to the case of our multigraph $G'$.
        \end{enumerate}
\end{enumerate}

The procedure is initialized with the best upper bound computed by the heuristic algorithm described in Section~\ref{section:method}. In the computational results of Section~\ref{section:results}, we will denote with \lbuno\ (\lbdue) the minimum of the lower bounds computed at Step 2-(i) (Step 2-(ii)) over the set of subsets of \ft-level routes. Lower bound \lbdue\ is computed using the lower bounds provided by the method of \cite{Baldacci2009}.

In the following, we describe how we extended the $ng$-routes relaxation to graph $G'$ and, for the sake of space, we omit the details of the procedure used to generate elementary routes, which is indeed a straightforward adaptation of the procedure used by \cite{Baldacci2009}.\\

\noindent
\textbf{Pricing ng-routes.}
The computation of the lower bounds at steps 1, 2-(i) and 2-(ii) and  the procedure used to generate elementary routes rely on the use of the $ng$-routes relaxation. In this section, we describe the extension of the relaxation described in \citet{Baldacci2011} to multigraph $G'$. We describe the relaxation for a generic satellite $k \in \Ns$ that, for sake of notation, is denoted with the index 0 in the description reported below.

Let $\Omega(w,j,i,p)$ be the subset of battery consumption values from vertex $j$ to arrive at vertex $i$ with a consumption equal to $w$, with $w \leq \ltot$, when $j$ is visited immediately before $i$ using arc of index $p$ of arc $(j,i) \in A'$. Set $\Omega(w,j,i,p)$ is defined as follows:
\begin{equation}
\Omega(w,j,i,p) =
\begin{cases}
\{w-c_{ji}\} & \text{if } s(j,i,p)=0  \text{ and } c_{ji} \leq w \\
\{w': 0 \leq w'+c_{jk} \leq \ltot\} & \text{if } s(j,i,p) = k \neq 0 \text{ and } c_{ki}=w  \\
\varnothing & \text{otherwise.} \\
\end{cases}
\end{equation}

Let $N_i \subseteq \Nc$ be a set of selected customers for vertex $i$ such that $N_i \ni i$ and $\vert N_i \vert \leq \deltap{N_i}$ ($\deltap{N_i}$ is an a priori defined parameter). The sets $N_i$ allow us to associate with each forward path $P=(0,i_1,\dots,i_k)$ in $G'$ the subset $\Pi(P) \subseteq V(P)$, $V(P)=\{0,i_1,\ldots,i_{k-1},i_k\}$, containing customer $i_k$ and every customer  $i_r$, $r=1,..,k-1$, of $P$ that belongs to all sets  $N_{i_{r+1}},\dots,N_{i_k}$  associated with the customers  $i_{r+1},\dots,i_k$  visited after $i_r$.
The set $\Pi(P)$ is defined as:
    $\Pi(P) = \{ i_r: i_r \in \bigcap_{s=r+1}^k N_{i_s}, r=1,\dots,k-1\} \cup \{i_k\}$.

\sloppy A \textit{$ng$-path} $(NG,q,w,i)$ is a non-necessarily elementary path $P=(0,i_1,\dots,i_{k-1},i_k=i)$ starting from the satellite 0 with an initial consumption equal to 0, visiting a subset of customers (even more than once) of total demand equal to $q$ such that $NG=\Pi(P)$, ending at customer $i$ with a total consumption equal to $w$, and such that $i \notin \Pi(P')$, where $P'=(0,i_1,\dots,i_{k-1})$ is an $ng$-path. We denote by $f(NG,q,w,i)$ the cost of the least cost $ng$-path $(NG,q,w,i)$.
An $(NG,q,w,i)$-route is an $(NG,q,w,0)$-path where $i$ is the last customer visited before arriving at the satellite.

Functions $f(NG,q,w,i)$ can be computed using \ac{DP}.
The state space graph $\lc{H}=(\lc{E},\Psi)$ is defined as follows:
$\lc{E} = \{(NG,q,w,i): q_i \leq q \leq \capvs, \forall NG \subseteq N_i \text{ s.t. } NG \ni i, \, \sum_{j \in NG}q_j \leq \capvs, \allowbreak \forall i \in \{0\} \cup \Nc, \forall w, 0 \leq w \leq \ltot \}$,
$\Psi=\{((NG',q',w',j),(NG,q,w,i))^p: \forall (NG',q',w',j) \in \Psi^{-1}(NG,q,w,j,i,p), \allowbreak
p=1,\dots,h(j,i), \forall (j,i) \in A', \forall (NG,q,w,i) \in \lc{E} \}$,
where $\Psi^{-1}(NG,q,w,j,i,p)=\{(NG',q-q_i,w',j): \forall NG' \subseteq N_j \text{ s.t. }NG' \ni j \text{ and } NG' \cap N_i={NG \setminus\{i\}}, \, \forall w' \in \Omega(w,j,i,p)\}$.

The \ac{DP} recursion for computing $f(NG,q,w,i)$ is:
\begin{equation}
f(NG,q,w,i) =  \hspace*{-0.5cm} \min_{\substack{ (j,i) \in A', \; 1 \leq p \leq h(j,i) \\ (NG',q',w',j) \in \Psi^{-1}(NG,q,w,j,i,p)}} \hspace*{-0.5cm}   \left\{f(NG',q',w',j)+d(j,i,p)\right\}, \forall (NG,q,w,i) \in \lc{E}, \label{ng.a11}
\vspace*{0.4cm}
\end{equation}
using as initial state $f(\{0\},0,0,0)=0$ and $f(\{0\},q,w,0)= \infty$ for $q > 0$ and $w > 0$.
In the computational experiments (Section \ref{section:results}), we set $\deltap{N_i}=12$, $\forall i \in \Nc$, and $N_i$ contains~$i$ and the 11 nearest customers to $i$.

From the experimental analysis presented in Section~\ref{section:results}, we observed that this mathematical programming algorithm can solve small and medium size instances to optimality and provide good lower bounds otherwise. Yet, this method requires a good initial upper bound to be truly effective, especially when the problem size grows. To produce these upper bounds, the following section introduces a metaheuristic based on large neighborhood search.

\section{Large Neighborhood Search}
\label{section:method}

Our metaheuristic, called LNS-E2E, follows the basic principles of ruin and recreate \citep{shaw1998using}. At each iteration, some parts of the solution are destroyed by a selected destroy operator (Section~\ref{sec:destroy}), and then repaired again (Section~\ref{sec:repair}) with a three-steps repair operator which reconstructs, in turn, the \sd-level routes, the \ft-level routes, and completes the reconstruction with an optimal insertion of visits to charging stations. Subsequently, a sophisticated local search (Section~\ref{sec:localsearch}) is applied to improve the resulting solution. During the local search, the labeling algorithm is used in combination with the moves to evaluate their impact.

The general structure of the method is presented in Algorithm~\ref{a_pseudoLNS}.
The sequence of destruction, reconstructions and local search is repeated until $i_{max}$ iterations have been performed without improvement of the incumbent solution (Lines 4--8). Once this termination criterion is attained, the best solution is stored (Lines 9--10) and the method performs a restart from a new random initial solution. This process repeats until a maximum time $T_\textsc{max}$ is attained (Lines 2--10).

	\begin{algorithm}[htbp]
      \linespread{1.1}\selectfont
	\DontPrintSemicolon
	\caption{LNS-E2E}   \label{a_pseudoLNS}
		$\mathcal{S}^{best} \leftarrow \varnothing$ \label{Alg:init}
		
		\While{\emph{CPU time} $< T_\textsc{max}$}{
		  $\mathcal{S} \leftarrow \text{LocalSearch(Repair(}\varnothing))$  \label{Alg:restart}			\tcc*{(re-)start: new solution}
			\For{$i \leftarrow 0$ \KwTo $i_{max}$}{
				$\mathcal{S}^{temp} \leftarrow \text{LocalSearch(Repair(Destroy(}\mathcal{S})))$\; \label{Alg:destroy}
				\If{$\emph{Cost}(\mathcal{S}^{temp}) < \emph{Cost}(\mathcal{S})$}{
					$\mathcal{S} \leftarrow \mathcal{S}^{temp}$ 	\label{Alg:newIncumb}						\tcc*{accept better solution}
					$i \leftarrow 0$ 
				}
			}
			\If{$\emph{Cost}(\mathcal{S}) < \emph{Cost}(\mathcal{S}^{best})$}{
				$\mathcal{S}^{best} \leftarrow \mathcal{S}$ 																		\tcc*{store best solution}
			}
		} 
		\Return $\mathcal{S}^{best}$\;
	\end{algorithm}

In contrast with the adaptive large neighborhood search of \cite{Pisinger2007}, LNS-E2E makes uses of a very limited number of destroy operators, and a single repair operator. Moreover, the probabilities of use of each operator are fixed, i.e., the method does not rely on adaptive mechanisms. This design is in line with the study of \cite{Breunig2016}, where it was observed that the algorithm with a simple fixed probability selection equaled its adaptive counterpart on the 2EVRP. The following subsections now describe each component of the method in deeper details.

\subsection{Destroy operators}
\label{sec:destroy}

At each iteration, one out of three destroy operators is selected with equal probability:
\begin{itemize}[nosep,leftmargin=*]
\item
\textbf{A) Related nodes removal.} A seed customer is randomly chosen. A random number of its Euclidean closest customers as well as the seed customer are removed from the current solution and added to the list of nodes to re-insert. This operator receives a parameter $p_1$, which denotes the maximum percentage of nodes to remove. At most $\lceil p_1 \cdot n_c \rceil$ nodes are removed.
\item
\textbf{B) Random routes removal.} Randomly selects routes and removes the associated customers visited, adding them to the list of nodes to re-insert. This operator randomly selects a number of routes in the interval $[0,\lceil p_2 \cdot \frac{q_\textsc{tot}}{Q_2} \rceil]$. The last term gives a lower bound on the number of routes needed to serve all customers.
\item
\textbf{C) Close satellite.} Chooses a random satellite. If the satellite can be closed and the remaining open ones still can provide sufficient capacity for a feasible solution, the chosen satellite is closed temporarily. All customers that are assigned to it are removed and added to the list of nodes to re-insert. The satellite stays closed until it is opened again in a later phase.\\
\end{itemize}

Moreover, the following two other operators may be applied right after one of the destroy operators described above:
\begin{itemize}[nosep,leftmargin=*]
\item
\textbf{D) Open all satellites.} With a probability of $\hat{p}_3$, all currently closed satellites are set to be available again in future repair phases.
\item
\textbf{E) Remove single customer routes.} This operator removes all routes which contain only one single customer. Typically, a complete solution does not often contain any route matching this criterion, but this can happen after a partial destruction. Therefore, with a probability of $\hat{p}_4$, all those customers which remain on a single node route after the destruction phase are also added to the list to re-insert. As there is a limit on the number of vehicles available, removing short routes also allows to use a vehicle originating from another satellite in the next repair phase.
\end{itemize}

\subsection{Repair operator and initial solution construction}
\label{sec:repair}

We propose a repair operator based on three steps, which first reinserts customer-visits in \sd-level routes, then reconstructs \ft-level routes, and finally completes the solution with recharging stations visits. Note also that the creation of the initial solution can be seen as a totally destroyed or empty solution ($\varnothing$), and therefore follows the same principle.\\

\noindent
\textbf{Reinsertion of customer visits.}
The classic cheapest insertion calculates every possible insertion position for every node to insert and selects the least-cost one. LNS-E2E uses a simplified version of this greedy heuristic with lower complexity, which iteratively inserts the first node from the insertion list in its cheapest position, until all nodes have been inserted. As a consequence, the outcome of the reconstruction depends on the order of the nodes in the list, favoring diversification.

The order of nodes in the list is randomly shuffled prior to insertions. In the exceptional case where this method fails to generate a feasible solution, another construction is attempted, this time ordering the nodes in the list by decreasing demand quantity. Such an ordering has a better chance to result in a feasible solution with respect to the capacity (i.e., packing) constraints, since no split deliveries are allowed on the second level. Overall, this first phase of the \sd-level routes reconstruction respects all constraints of the problem except those related to charging levels and recharging stations visits. \\

\noindent
\textbf{Construction of first level tours.}
After itineraries for the \sd-level routes have been found, the quantities needed at the satellites are known. With this information, the \ft-level routes can be reconstructed. We opted for a complete reconstruction, as the number of satellites is usually small and the \sd-level routes can very significantly change from one iteration to another.  On the first level, split deliveries are not only allowed, but sometimes also necessary to find a feasible solution. Depending on the customers associated to a satellite, it can occur that the requested quantity at the satellite is larger than a full truckload. Therefore, we use a simple preprocessing step: for any satellite with a demand larger than a full truckload, we create a back-and-forth trip from the depot, until the remaining demand is smaller than a truck's capacity. The simplified cheapest insertion is then used to complete the \ft-level solution. In practical settings with up to 10 or 20 satellite facilities, this method finds optimal \ft-level routes in a majority of cases in a very limited computational effort.\\

\begin{figure}[htbp]
\centering
\includegraphics[width = 0.8\textwidth]{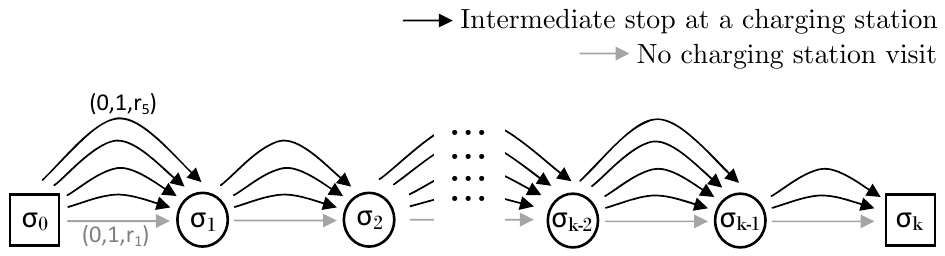}
\caption{Illustration of the multigraph $\cH$. Non-dominated choices of charging stations visits are represented by parallel arcs.}
\label{DP-Charging}
\end{figure}

\noindent
\textbf{Optimal insertion of charging stations visits.}
At this point, the algorithm has reconstructed a solution which is feasible in terms of load capacities but usually infeasible in terms of battery capacities. To restore feasibility, it uses a \ac{DP} algorithm which finds the optimal charging stations positions for each \sd-level route. The problem of inserting charging stations visits in a route $\boldsymbol\sigma = (\sigma_0,\sigma_1,\dots,\sigma_{K})$ can be reduced to a \ac{SPPRC} in a directed acyclic multigraph $\cH=(\cN,\cA)$, such that $\cN = \{0\} \cup \{1,\dots,K-1\} \cup \{K\}$. The nodes $0$ and $K$ correspond to depot visits (such that $\sigma_0 = \sigma_{K} = 0$), while the other nodes represent customer visits. Each arc $(i,i+1,r_k) \in \cA$ corresponds to a non-dominated arc between $\sigma_i$ and $\sigma_{i+1}$, with the same characteristics as $(\sigma_i,\sigma_{i+1},r_k) \in A'$ defined in Section \ref{section:description} and possible en-route recharging. This multigraph is illustrated in Figure~\ref{DP-Charging}.

Solving this \ac{SPPRC} can be simply done using Bellman's algorithm in the topological order $(0,1,\dots,K)$. Using the same notations as in previous sections, the state space graph $\lc{\bar{H}}=(\lc{\bar{E}},\bar{\Psi})$ is defined as $\lc{\bar{E}} = \{(w,i): \forall w, 0 \leq w \leq \ltot , \forall i \in \cN \}$ and
$\bar{\Psi}=\{((w',i-1),(w,i))^p: \forall (w',{i-1}), w' \in \Omega(w,\sigma_{i-1},\sigma_{i},p),
p=1,\dots,h(\sigma_{i-1},\sigma_{i}), \forall (w,i) \in \lc{\bar{E}} \}$.
Defining $f(w,i)$ as the minimum cost of a path starting from $0$ and reaching node $i$ with battery consumption $w$, the \ac{DP} recursion can be expressed as:
\begin{equation}
f(w,i) = \min_{\substack{1 \leq p \leq h(\sigma_{i-1},\sigma_i) \\w' \in \Omega(w,\sigma_{i-1},\sigma_{i},p)}}\{f(w',i-1)+d(\sigma_{i-1},\sigma_i,p)\}, \forall (w,i) \in \lc{\bar{E}},
\vspace*{0.4cm}
\end{equation}
and the initial state is set as $f(0,0)=0$ and $f(w,0)= \infty$ for $w > 0$.

In the rare case where no feasible path can be found at the end of the \ac{DP} recursion, a second execution of the \ac{DP} algorithm is done, with a minor modification of the label propagation function allowing to use and penalize battery capacity excesses. In this case, any consumption over the battery level $w > L$ is converted into a proportional penalty of $M \times (w-L)$, where $M$ is a large constant. Therefore, the method seeks a route with the smallest penalty in priority, and then the shortest distance. Due to its large impact on the objective, this infeasibility will generally be resolved in the next steps of the method: the local search or the next destroy and repair phase.

\subsection{Local search with systematic charging stations relocations}
\label{sec:localsearch}

After solution reconstruction, LNS-E2E applies a local search procedure on the \sd-level routes
based on \textsc{2-opt}, \textsc{2-opt*}, \textsc{Relocate}, \textsc{Swap} and \textsc{Swap2-1} moves \citep[see, e.g.,][for a detailed description of these neighborhoods]{Vidal2013}. The \textsc{2-opt*} moves are only tested between routes originated from the same satellite. Moreover, when testing moves that involve routes from different satellites, the algorithm checks that enough capacity is available in the satellites and the associated \ft-level routes.
The moves are tested in random order and a first-improvement acceptance policy is used, i.e., any move which results in an improvement in terms of cost is directly applied, until no more improvement can be found.
Similarly to the granular search by \cite{Toth2003}, the moves are limited to node pairs $(i,j)$ such that $j$ belongs to one of the $\Gamma$ closest vertices from $i$.

Most modification of the sequence of customer visits or their assignment to vehicles induce some necessary changes in the planning of charging stations visits. Ideally, one would like to apply the labeling algorithm described in Section \ref{sec:repair} to obtain the \emph{exact} cost of each move, with an optimal placement of recharging stations in each newly-created route. Such an evaluation would be, however, prohibitive in terms of computational effort. To speed up the method with only a minimal impact on solution quality, we propose some heuristic move filters, which are quite similar in principle to the techniques used by \cite{Taillard1997} for the VRP with soft time windows. We first evaluate each move without the labeling algorithm to obtain an approximation of its impact on the total distance. When doing this calculation, the current locations of the charging stations are unchanged. Any move which is feasible in terms of load capacity and does not deteriorate the total distance by more than $3\%$ is then evaluated exactly in combination with the labeling algorithm, so as to find better charging stations locations which may lead to an improvement. After this exact evaluation, any improving move is applied.

\section{Computational Experiments}
\label{section:results}

We conducted extensive experimental analyses with two aims. Firstly, we evaluate the performance of the proposed algorithms for different types of instances, and measure the benefit of integrated routing and recharging-stations planning  (Section~\ref{section:analysis1}). For this analysis, we extend classical \ac{2E-VRP} instances into \ac{E2E-VRP} instances in order to obtain diverse and challenging datasets and allow possible comparisons with previous algorithms. Secondly, we analyze the impact of two defining features of electric-fueled city-distribution networks: the density of charging stations in a city, and the vehicles' battery capacities (Section~\ref{section:analysis3}). For this analysis, we produced a new set of medium-scale instances which simulates a realistic delivery scenario in a metropolitan area, using battery specifications from recent electric vehicles.

The mathematical programming algorithm was coded in Fortran 77, and run on a single thread of a 3.6 GHz Intel i7-4790 CPU with 32GB of RAM. It relies on CPLEX 12.5.1 for the resolution of the linear programs and some integer subproblems. The metaheuristic was coded in Java (JRE 1.8.0--151), and run on a single thread of a 3.4 GHz Intel i7-3770 CPU. All benchmark instances used in this paper are available for download at \url{https://www.univie.ac.at/prolog/research/electric2EVRP} and \url{https://w1.cirrelt.ca/~vidalt/en/VRP-resources.html}.

\subsection{Method performance and benefits of integrated planning}
\label{section:analysis1}

Our benchmark instances for this first analysis are natural extensions of the \ac{2E-VRP} instances known as Set~2 and 3 by \cite{Perboli2011}, Set~5 by \cite{Hemmelmayr2012}, and Set~6 by \cite{Baldacci2013}. The depot, satellite and customers locations remain unchanged. The new information is associated to the electric vehicles (maximum charging level and energy consumption) and to the charging stations (coordinates).

The selection of charging stations follows the guidelines of \cite{Schneider2013} (instances of the \ac{EVRPTWRS}) and \cite{Desaulniers2016}. The ratio between charging stations and customers is chosen between $1/10$ and $1/5$. Firstly, every depot and satellite location provides charging abilities vehicles. To pick the remaining locations, we defined a grid of $100 \times 100$ candidate locations based upon the range of x- and y-axis coordinates from the existing \ac{2E-VRP} instances. For each location, we counted the number of customers in ``close proximity'', defined as half the average tour length in the best known \ac{2E-VRP} solution. The more customers one candidate location has in proximity, the more likely it is selected as a charging station. This was achieved by a roulette wheel selection of the remaining charging stations among those 10,000 locations. Finally, all distances are calculated as Euclidean and rounded to the nearest integer value. To reduce the effect of rounding, all x- and y-coordinates from the classical \ac{2E-VRP} instances have been multiplied by a factor ten.
For each instance, the battery capacity has been defined as $L=\max \{0.6\,\gamma_1, 2.0\, \gamma_2\}$, where $\gamma_1$ represents the average route length of all second-level routes in the best-known \ac{2E-VRP} solution, and $\gamma_2$ is the largest distance of a customer to its closest recharging station. For the sake of simplicity, the energy consumption per distance unit is always set to 1 ($d_{ij} = c_{ij}$). As highlighted in our computational experiments, this approach leads to feasible solutions for all the instances, whereas the best-known \ac{2E-VRP} solutions are generally not feasible for the corresponding~\ac{E2E-VRP}~instances.\\

\noindent
\textbf{Parameters calibration.}
To produce suitable values for the new parameters of the LNS, we used a preliminary meta-calibration based on the \ac{CMA-ES} of \cite{hansen2006}. During meta-calibration, the parameters are considered to be the decision variables, and the associated objective corresponds to the average solution quality of LNS-E2E over ten runs on a set of training instances.
This training set includes six larger-scale instances from Set~5: \{100-5-1, 100-5-2b, 100-10-1, 100-10-2b, 200-10-1, 200-10-2b\}.
Table~\ref{tab:parametertuning} lists the method parameters, the allowed range for each parameter, and the final values found by the meta-calibration process. These values will be used for the rest of the experiments.

After calibration, we evaluated the proposed mathematical programming algorithm and the metaheuristic on the complete set of benchmark instances. The termination criterion of LNS-E2E has been set to $T_\textsc{max} = 150$ seconds for the small instances of Sets 2, 3 and 6a, and 900 seconds for the large-scale instances of Set 5.

\begin{table}[htbp]
 \centering
 \caption{Range of parameters used during meta-calibration, and final values found}
 \begin{tabular}{rlrr}
 \toprule
 & Parameter & Search interval & Final value \\
 \midrule
 $p_1$ & Related removal (\%) & 0--100 & 11 \\
 $p_2$ & Random route removal (\%) & 0--100 & 37 \\
 $\hat{p}_3$ & Open all satellites (\%) & 0--100 & 12 \\
 $\hat{p}_4$ & Remove single customer routes (\%) & 0--100 & 18 \\
 $\tau$ & Granularity threshold for move evaluations & 0--40 & 25 \\
 $i_{max}$ &Number of non-improving iterations before restart & 0--1000 & 385 \\
 \bottomrule
 \end{tabular}%
 \label{tab:parametertuning}%
\end{table}%

\noindent
\textbf{Results on small instances.} Table~\ref{tab:n22} reports the results on the smaller instances of Set~2 and Set~3 with 21 customers. The leftmost group of columns reports the characteristics of the instances: number of customers \emph{$n_c$}, satellites~\emph{$n_s$}, trucks~\emph{$m^1$}, (electric) second level vehicles \emph{$m^2$} and charging stations \emph{$n_r$}.
The next group of columns shows the performance of the mathematical programming algorithm. The column UB reports the best upper bound at the end of the algorithm, and the solutions marked with an asterisk are proven optimal. The lower bounds obtained at different steps of the exact method are also displayed along with the associated CPU time values, using the same notations as described in Section~\ref{sec:exm}.
The next group of columns reports the performance of LNS-E2E: the average (Avg) and best (Best) solution quality over five runs, the average computational time per run (T(s)), and the average time per run needed to reach the final solution of the run (T*(s)).
The overall best known solution (BKS) found during all experiments (including calibration and testing) is reported in the rightmost column.

The exact algorithm produced optimal solutions (marked with an asterisk) for all instances except one. A notable improvement is visible when comparing LB2, obtained by repeated resolutions of \ac{MDC-VRP} with side constraints, with LB0 and LB1. The CPU time of the method remains below one minute for 4/12 instances, but can rise up to six hours in other cases, illustrating the difficulty of the \ac{E2E-VRP}, as the presence of the battery capacity constraints significantly increases the time needed for route enumeration. For these instances, the metaheuristic always found the optimal solutions in at least one run out of five. Still, we observe some variance in the results of different independent runs. This is due to the combination of multiple classes of decision variables (two-echelon routing, satellite selection and charging stations selection), which make the problem very intricate and favors the creation of many local minima. The LNS-E2E remains nonetheless accurate, with an average gap of 1.18\% from the optimal or best known solutions. The time taken to attain the final solution of the run varies from 2 to 132 seconds, depending on the instance.\\

\noindent
\textbf{Results on medium instances.}
Table~\ref{tab:lowerbounds} displays the results on the medium instances of Sets 2, 3 and 6a, containing between 32 to 75 customers. The same convention as the previous table is used.
For this scale of instances, the mathematical programming algorithm does not generate proven optimal solutions in the allotted time and was stopped after the computation of bound LB1. The average optimality gap between the best solution found by LNS-E2E and the bound LB1, for this group of instances, amounts to 3.57\%, demonstrating the good accuracy of both approaches.
As usual when comparing exact algorithms with metaheuristics, the difference of CPU time between the two methods becomes more marked for larger instances. For some instances with 75 customers, the time needed to compute LB1 grows as high as 25 hours, whereas the termination of the heuristic is guaranteed after 150 seconds.\\

\noindent
\textbf{Results on large instances, and impact of integrated routing and recharging stations optimization.}
The larger instances of Set 5 contain 100 or 200 customer visits. To the best of our knowledge, only 3/18 instances have been solved to proven optimality for the classical \ac{2E-VRP} (without considering electric vehicles and recharging stations). The \ac{E2E-VRP} appears to be even harder to solve, and our exact approach could not produce optimal solutions or good lower bounds in reasonable time.
For this set of instances, we therefore concentrate our analysis on the results of the metaheuristic, with the aim of assessing the performance of the algorithm and the impact of an integrated optimization of routing and recharging stations decisions.
To that end, we compared four algorithms.
The first two algorithms solve the \ac{2E-VRP} \emph{without} electric vehicles:
\begin{itemize}[nosep,leftmargin=*]
\item \textbf{LNS-2E}: the algorithm presented in \cite{Breunig2016} (LNS-2E), which produces the current state-of-the-art results for that problem;
\item \textbf{LNS-E2E $\boldsymbol\infty$}: the proposed algorithm, in which the range of the electric vehicles is set to $\infty$ to make recharging-stations visits unnecessary.
\end{itemize}
The two other solution methods are designed for the problem \emph{with} electric vehicles:
\begin{itemize}[nosep,leftmargin=*]
\item \textbf{LNS-2E post}: resolution of the classical \ac{2E-VRP} (disregarding battery constraints) with LNS-2E, followed by a post-optimization using the labeling algorithm to insert charging-stations visits;
\item \textbf{LNS-E2E}:  the proposed algorithm, with integrated routing and planning of charging stations.
\end{itemize}
Each method was run until a time limit of 15 minutes, and the same rounding convention (integer distances) have been adopted to allow direct solution comparisons.
Table~\ref{tab:set5} reports the average (Avg) and best (Best) solution quality of each method over ten runs, as well as the average CPU time to reach the final solution of each run (T*(s)). For future reference, the BKS found on Set~5 for the LNS-E2E during preliminary calibration and testing are also listed in the rightmost column.

Firstly, these results highlight the good accuracy of the proposed LNS-E2E, even for the particular case of the \ac{2E-VRP} \emph{without} electric vehicles. In comparison with the current state-of-the-art algorithm LNS-2E, better average quality solutions are found on all 200-customer instances, with improvements rising up to 2.41\%, while solutions of slightly lower quality are obtained on the 100-customer test cases.

Secondly, we observe the large benefits of an integrated routing and charging stations visits planning. Even when starting with good \ac{2E-VRP} solutions, a post-ex insertion of charging stations results in solutions of poor quality for the \ac{E2E-VRP} in comparison with the integrated LNS-E2E approach. The average gain related to an integrated optimization in comparison to post-optimization amounts to 3.28\%, and can reach as high as 7.93\% for instance 100-10-2b. Finally, in terms of computational effort, we observe that the proposed LNS-E2E approach finds solutions in a similar time as LNS-2E, despite the joint optimization of charging stations. This is essentially due to the heuristic move filters described in Section \ref{sec:localsearch}, allowing to evaluate and discard a large proportion of non-promising local search moves without a call to the labeling algorithm. The next section will study in deeper details the impact of some of these method components.\\

\begin{landscape}
\vspace*{2cm}
\begin{table}[htbp]
	\tiny
\renewcommand{\arraystretch}{1.2}
\setlength{\tabcolsep}{3pt}
	\centering
  \caption{Performance analysis on small instances of Set 2 and 3\label{tab:n22}}
		\begin{tabular}{lrrrrrr@{\hspace*{0cm}}rrrrrrrrrrrrrr}
    \toprule
       &   \multicolumn{5}{c}{Characteristics} &  \multicolumn{10}{c}{Exact Method} & \multicolumn{5}{c}{LNS-E2E} \\
    \cmidrule(lr){2-6} \cmidrule(lr){7-16} \cmidrule(lr){17-21}
    \multicolumn{1}{c}{Instance} & \multicolumn{1}{c}{$n_c$} & \multicolumn{1}{c}{$n_s$} & \multicolumn{1}{c}{$m^1$} & \multicolumn{1}{c}{$m^2$} & \multicolumn{1}{c}{$n_r$} & \multicolumn{1}{c}{UB} & \multicolumn{1}{c}{} & \multicolumn{1}{c}{\%LB0} & \multicolumn{1}{c}{LB0} & \multicolumn{1}{c}{T$_\text{LB0}$(s)} & \multicolumn{1}{c}{\%LB1} & \multicolumn{1}{c}{LB1} & \multicolumn{1}{c}{\%LB2} & \multicolumn{1}{c}{LB2} & \multicolumn{1}{c}{T$_\text{All}$(s)} & \multicolumn{1}{c}{Avg} & \multicolumn{1}{c}{Best}  & \multicolumn{1}{c}{T(s)} & \multicolumn{1}{c}{T*(s)} & \multicolumn{1}{c}{BKS}\\
		\midrule
    \multicolumn{21}{l}{\textbf{Set2}} \\
    n22-k4-s6-17 & 21    & 2     & 3     & 4     & 4     & 5229  & *     & 98.23\% & 5136.3 & 2.3   & 98.87\% & 5169.9 & 100.00\% & 5229.00 & 5.6   & \textbf{5229.0} & \textbf{5229}  & 150  & 6.1 & \textbf{5229}\\
    n22-k4-s8-14 & 21    & 2     & 3     & 4     & 4     & 5094  & *     & 95.95\% & 4887.6 & 2.9   & 95.95\% & 4887.6 & 100.00\% & 5094.00 & 77.2  & 5168.4 & \textbf{5094}  & 150   & 39.8 & \textbf{5094}\\
    n22-k4-s9-19 & 21    & 2     & 3     & 4     & 4     & 5236  & *     & 93.95\% & 4919.1 & 4.3   & 93.95\% & 4919.1 & 99.62\% & 5216.26 & 14.3  & 5240.2 & \textbf{5236}  & 150   & 65.3 & \textbf{5236}\\
    n22-k4-s10-14 & 21    & 2     & 3     & 4     & 4     & 5561  & *     & 96.99\% & 5393.6 & 4.2   & 96.99\% & 5393.6 & 99.64\% & 5541.15 & 343.9 & \textbf{5561.0} & \textbf{5561}  & 150   & 2.0 & \textbf{5561}\\
    n22-k4-s11-12 & 21    & 2     & 3     & 4     & 4     & 5793  & *     & 95.92\% & 5556.6 & 3.6   & 95.92\% & 5556.6 & 99.24\% & 5748.74 & 34.1  & 5822.0 & \textbf{5793}  & 150   & 74.8 & \textbf{5793}\\
    n22-k4-s12-16 & 21    & 2     & 3     & 4     & 4     & 4125  & *     & 96.96\% & 3999.5 & 5.6   & 97.17\% & 4008.1 & 100.00\% & 4125.00 & 8.7   & 4211.4 & \textbf{4125}  & 150   & 77.5 & \textbf{4125}\\
    \addlinespace[5pt]
		Average &       &       &       &       &       & 5173.0 &       & 96.33\% & 4982.1 & 3.8   & 96.47\% & 4989.2 & 99.75\% & 5159.02 & 80.6  & 5205.3 & \textbf{5173.0}  & 150   & 44.3 & \textbf{5173.0}\\
		\midrule
    \multicolumn{21}{l}{\textbf{Set3}} \\
    n22-k4-s13-14 & 21    & 2     & 3     & 4     & 4     & 6396  & *     & 95.95\% & 6137.2 & 3.8   & 95.95\% & 6137.2 & 99.77\% & 6381.23 & 795.2 & 6406.8 & \textbf{6396}  & 150   & 15.8 & \textbf{6396}\\
    n22-k4-s13-16 & 21    & 2     & 3     & 4     & 4     & 6922  & *     & 97.00\% & 6714.0 & 2.1   & 97.00\% & 6714.0 & 99.86\% & 6912.52 & 515.1 & 6954.2 & \textbf{6922}  & 150   & 31.5 & \textbf{6922}\\
    n22-k4-s13-17 & 21    & 2     & 3     & 4     & 4     & 6408  & *     & 93.57\% & 5996.2 & 4.1   & 93.57\% & 5996.2 & 98.62\% & 6319.83 & 421.1 & \textbf{6408.0} & \textbf{6408}  & 150   & 35.4 & \textbf{6408}\\
    n22-k4-s14-19 & 21    & 2     & 3     & 4     & 4     & 6634  &       & 95.27\% & 6320.4 & 2.9   & 95.27\% & 6320.4 & 98.60\% & 6541.32 & 11995.6 & 7018.4 & \textbf{6634}  & 150   & 132.1 & \textbf{6634}\\
    n22-k4-s17-19 & 21    & 2     & 3     & 4     & 4     & 6947  & *     & 95.79\% & 6654.2 & 4.2   & 95.79\% & 6654.2 & 98.24\% & 6824.66 & 20575.4 & 7094.6 & 6965    & 150   & 72.4 & 6965\\
    n22-k4-s19-21 & 21    & 2     & 3     & 4     & 4     & 6529  & *     & 96.60\% & 6307.3 & 8.2   & 96.67\% & 6311.6 & 99.47\% & 6494.46 & 8499.0 & 6625.2 & \textbf{6529}  & 150   & 104.9 & \textbf{6529}\\
    \addlinespace[5pt]
		Average &       &       &       &       &       & 6639.3 &       & 95.70\% & 6354.9 & 4.2   & 95.71\% & 6355.6 & 99.09\% & 6579.00 & 7133.6 & 6751.2 & 6642.3  & 150   & 65.4 & 6642.3\\
    
    \bottomrule
    \end{tabular}%
\end{table}
\end{landscape}
\begin{center}
\begin{table}[htbp]
  \tiny
	\tabcolsep=2pt
       \renewcommand{\arraystretch}{1.2}
  \centering
   \vspace*{1cm}
   \caption{Performance analysis on medium-scale instances of Set 2, 3 and 6a \label{tab:lowerbounds}}
    \begin{tabular}{lrrrrrrrrrrrrrrrrr}
    \toprule
          & \multicolumn{5}{c}{Characteristics} & \multicolumn{7}{c}{Lower Bounds} & \multicolumn{5}{c}{LNS-E2E}            \\
    \cmidrule(lr){2-6} \cmidrule(lr){7-13} \cmidrule(lr){14-18}
    \multicolumn{1}{c}{Instance} & \multicolumn{1}{c}{$n_c$} & \multicolumn{1}{c}{$n_s$} & \multicolumn{1}{c}{$m^1$} & \multicolumn{1}{c}{$m^2$} & \multicolumn{1}{c}{$n_r$} & \multicolumn{1}{c}{UB} & \multicolumn{1}{c}{\%LB0} & \multicolumn{1}{c}{LB0} & \multicolumn{1}{c}{T$_\text{LB0}$(s)} & \multicolumn{1}{c}{\%LB1} & \multicolumn{1}{c}{LB1} &  \multicolumn{1}{c}{T$_\text{LB1}$(s)} & \multicolumn{1}{c}{Avg} & \multicolumn{1}{c}{Best} & \multicolumn{1}{c}{BKS} & \multicolumn{1}{c}{T(s)} & \multicolumn{1}{c}{T*(s)} \\
		\midrule
    \multicolumn{18}{l}{\textbf{Set 2}} \\
    n33-k4-s1-9 & 32    & 2     & 3     & 4     & 5     & 7617  & 98.46\% & 7499.4 & 73.3  & 98.46\% & 7499.4 & 149.4 & 7751.0 & \textbf{7617} & \textbf{7617} & 150   & 75.3 \\
    n33-k4-s2-13 & 32    & 2     & 3     & 4     & 5     & 7925  & 94.81\% & 7513.4 & 44.7  & 94.81\% & 7513.4 & 112.7 & 8025.0 & \textbf{7925} & \textbf{7925} & 150   & 103.3 \\
    n33-k4-s3-17 & 32    & 2     & 3     & 4     & 5     & 8090  & 92.88\% & 7514.2 & 69.7  & 92.88\% & 7514.2 & 181.3 & 8280.2 & \textbf{8090} & \textbf{8090} & 150   & 107.0 \\
    n33-k4-s4-5 & 32    & 2     & 3     & 4     & 5     & 8870  & 93.84\% & 8323.8 & 79.2  & 93.84\% & 8323.8 & 257.1 & 8925.2 & \textbf{8870} & \textbf{8870} & 150   & 91.0 \\
    n33-k4-s7-25 & 32    & 2     & 3     & 4     & 5     & 8318  & 95.51\% & 7944.1 & 77.0  & 95.74\% & 7963.3 & 168.9 & 8374.8 & \textbf{8318} & \textbf{8318} & 150   & 92.7 \\
    n33-k4-s14-22 & 32    & 2     & 3     & 4     & 5     & 8621  & 98.42\% & 8484.4 & 218.5 & 98.42\% & 8484.4 & 529.5 & 8680.4 & \textbf{8621} & \textbf{8621} & 150   & 90.2 \\
    \addlinespace[5pt]
		Average &       &       &       &       &       & 8240.2 & 95.65\% & 7879.9 & 93.7  & 95.69\% & 7883.1 & 233.1 & 8339.4 & \textbf{8240.2} & \textbf{8240.2} & 150   & 93.3 \\
		\midrule
    \multicolumn{18}{l}{\textbf{Set 3}}\\
    n33-k4-s16-22 & 32    & 2     & 3     & 4     & 6     & 7561  & 91.60\% & 6926.2 & 89.4  & 91.60\% & 6926.2 & 328.4 & 7656.2 & \textbf{7561} & \textbf{7561} & 150   & 94.5 \\
    n33-k4-s16-24 & 32    & 2     & 3     & 4     & 6     & 7501  & 94.77\% & 7108.5 & 168.4 & 94.77\% & 7108.8 & 607.1 & 7520.0 & \textbf{7501} & \textbf{7501} & 150   & 102.7 \\
    n33-k4-s19-26 & 32    & 2     & 3     & 4     & 6     & 7212  & 94.42\% & 6809.5 & 98.6  & 94.42\% & 6809.5 & 253.4 & 7223.2 & \textbf{7212} & \textbf{7212} & 150   & 47.3 \\
    n33-k4-s22-26 & 32    & 2     & 3     & 4     & 6     & 7334  & 95.81\% & 7027.0 & 290.5 & 96.85\% & 7103.1 & 738.5 & 7498.4 & \textbf{7334} & \textbf{7334} & 150   & 131.3 \\
    n33-k4-s24-28 & 32    & 2     & 3     & 4     & 6     & 7443  & 95.40\% & 7100.5 & 234.2 & 96.80\% & 7204.6 & 569.9 & 7371.6 & \textbf{7443} & \textbf{7443} & 150   & 116.2 \\
    n33-k4-s25-28 & 32    & 2     & 3     & 4     & 6     & 7429  & 93.68\% & 6959.7 & 258.4 & 93.68\% & 6959.7 & 579.8 & 7490.4 & \textbf{7429} & \textbf{7429} & 150   & 108.3 \\
    \addlinespace[5pt]
		Average &       &       &       &       &       & 7413.3 & 94.28\% & 6988.6 & 189.9 & 94.69\% & 7018.6 & 512.9 & 7460.0 & \textbf{7413.3} & \textbf{7413.3} & 150   & 100.0 \\
    \midrule
		\multicolumn{18}{l}{\textbf{Set 6a}} \\
    A-n51-4 & 50    & 4     & 2     & 50    & 5     & 7663  & 95.27\% & 7300.8 & 121.9 & 98.76\% & 7568.0 & 762.9 & 7879.4 & \textbf{7663} & \textbf{7663} & 150   & 109.4 \\
    A-n51-5 & 50    & 5     & 2     & 50    & 6     & 8268  & 95.77\% & 7918.0 & 98.4  & 98.16\% & 8116.0 & 2783.6 & 8386.4 & \textbf{8268} & \textbf{8268} & 150   & 64.5 \\
    A-n51-6 & 50    & 6     & 2     & 50    & 7     & 7795  & 93.08\% & 7255.9 & 117.1 & 98.18\% & 7653.4 & 15723.7 & 7943.8 & \textbf{7795} & \textbf{7795} & 150   & 106.0 \\
    A-n76-4 & 75    & 4     & 3     & 75    & 7     & 10599 & 95.40\% & 10111.7 & 214.9 & 97.23\% & 10305.4 & 6463.9 & 10692.0 & \textbf{10599} & \textbf{10599} & 150   & 97.0 \\
    A-n76-5 & 75    & 5     & 3     & 75    & 7     & 11178 & 95.18\% & 10638.9 & 175.5 & 98.17\% & 10973.6 & 16406.4 & 11242.4 & \textbf{11178} & \textbf{11178} & 150   & 88.8 \\
    A-n76-6 & 75    & 6     & 3     & 75    & 7     & 10156 & 95.60\% & 9709.2 & 242.2 & 98.92\% & 10046.1 & 85538.4 & 10250.0 & \textbf{10156} & \textbf{10156} & 150   & 110.7 \\
    B-n51-4 & 50    & 4     & 2     & 50    & 5     & 6589  & 97.20\% & 6404.4 & 163.2 & 97.96\% & 6454.8 & 342.4 & 6791.2 & \textbf{6589} & \textbf{6589} & 150   & 111.6 \\
    B-n51-5 & 50    & 5     & 2     & 50    & 6     & 7252  & 94.73\% & 6869.8 & 116.3 & 95.53\% & 6928.0 & 1859.6 & 7446.4 & \textbf{7252} & 7240 & 150   & 90.8 \\
    B-n51-6 & 50    & 6     & 2     & 50    & 7     & 6583  & 95.02\% & 6255.0 & 256.0 & 97.51\% & 6419.3 & 3054.1 & 6787.6 & \textbf{6583} & \textbf{6583} & 150   & 61.4 \\
    B-n76-4 & 75    & 4     & 3     & 75    & 7     & 9945  & 95.99\% & 9546.7 & 198.4 & 98.02\% & 9748.0 & 2184.4 & 9995.8 & \textbf{9945} & 9943 & 150   & 99.5 \\
    B-n76-5 & 75    & 5     & 3     & 75    & 7     & 9139  & 94.70\% & 8655.1 & 210.4 & 98.26\% & 8980.0 & 9903.7 & 9209.2 & \textbf{9139} & \textbf{9139} & 150   & 71.9 \\
    B-n76-6 & 75    & 6     & 3     & 75    & 7     & 8238  & 94.44\% & 7780.1 & 427.0 & 97.80\% & 8056.5 & 79962.3 & 8287.6 & \textbf{8238} & \textbf{8238} & 150   & 82.4 \\
    C-n51-4 & 50    & 4     & 2     & 50    & 5     & 8407  & 94.57\% & 7950.2 & 137.0 & 95.74\% & 8048.5 & 888.4 & 8596.2 & \textbf{8407} & \textbf{8407} & 150   & 80.4 \\
    C-n51-5 & 50    & 5     & 2     & 50    & 6     & 8810  & 94.99\% & 8368.3 & 261.1 & 95.77\% & 8437.3 & 1346.5 & 9276.0 & \textbf{8810} & \textbf{8810} & 150   & 82.4 \\
    C-n51-6 & 50    & 6     & 2     & 50    & 7     & 8160  & 93.73\% & 7648.6 & 180.3 & 95.83\% & 7819.7 & 7092.7 & 8390.6 & \textbf{8160} & \textbf{8160} & 150   & 72.9 \\
    C-n76-4 & 75    & 4     & 3     & 75    & 7     & 12162 & 94.61\% & 11506.7 & 199.0 & 98.30\% & 11955.7 & 3996.1 & 12381.2 & \textbf{12162} & 12147 & 150   & 99.3 \\
    C-n76-5 & 75    & 5     & 3     & 75    & 7     & 13033 & 92.00\% & 11990.5 & 402.2 & 93.33\% & 12163.4 & 38723.3 & 13247.0 & \textbf{13033} & \textbf{13033} & 150   & 79.3 \\
    C-n76-6 & 75    & 6     & 3     & 75    & 7     & 11808 & 93.28\% & 11014.5 & 285.0 & 97.11\% & 11466.6 & 93643.4 & 12129.8 & \textbf{11808} & 11806 & 150   & 92.8 \\
    \addlinespace[5pt]
		Average &       &       &       &       &       & 9210.3 & 94.75\% & 8718.0 & 211.4 & 97.25\% & 8952.2 & 20593.1 & 9385.1 & \textbf{9210.3} & 9208.6 & 150   & 88.9 \\
    \bottomrule
    \end{tabular}%
\end{table}%
\end{center}

\begin{landscape}
\vspace*{1cm}
\begin{table}[htbp]
	\tiny
	\tabcolsep=4pt
\renewcommand{\arraystretch}{1.2}
  \centering
  \caption{Performance analysis on the large-scale instances of Set~5 -- Evaluation of the benefits of an integrated routing and charging-stations optimization\label{tab:set5}}
    \begin{tabular}{lrrrrrrrrrrrrrrrrrr}
    \toprule
          &       &       &       &       &       & \multicolumn{6}{c}{2EVRP}                     & \multicolumn{7}{c}{E2EVRP}  \\
					 \cmidrule(lr){7-12} \cmidrule(lr){13-19}
          & \multicolumn{5}{c}{Characteristics}   & \multicolumn{3}{c}{LNS-2E} & \multicolumn{3}{c}{LNS-E2E $\infty$} & \multicolumn{3}{c}{LNS-2E post} & \multicolumn{3}{c}{LNS-E2E} &  \\
				\cmidrule(lr){2-6} \cmidrule(lr){7-9} \cmidrule(lr){10-12} \cmidrule(lr){13-15}	\cmidrule(lr){16-18}
    \multicolumn{1}{c}{Instance} & \multicolumn{1}{c}{$n_c$} & \multicolumn{1}{c}{$n_s$} & \multicolumn{1}{c}{$m^1$} & \multicolumn{1}{c}{$m^2$} & \multicolumn{1}{c}{$n_r$}    & \multicolumn{1}{c}{Avg} & \multicolumn{1}{c}{Best}& \multicolumn{1}{c}{T*(s)} & \multicolumn{1}{c}{Avg} & \multicolumn{1}{c}{Best} & \multicolumn{1}{c}{T*(s)} & \multicolumn{1}{c}{Avg} & \multicolumn{1}{c}{Best}& \multicolumn{1}{c}{T*(s)} & \multicolumn{1}{c}{Avg} & \multicolumn{1}{c}{Best} & \multicolumn{1}{c}{T*(s)} & \multicolumn{1}{c}{BKS} \\
				\midrule
    100-5-1 & 100   & 5     & 5     & 32    & 10    & 15692.6 & 15640 & 200.3 & 15660.9 & 15639 & 229.8 & 16689.9 & 16593 & 200.4 & 16224.6 & 16167 & 403.7 & 16165 \\
    100-5-1b & 100   & 5     & 5     & 15    & 10    & 11124.9 & 11082 & 405.9 & 11208.3 & 11118 & 215.7 & 12802.2 & 12495 & 406.0 & 12070.2 & 11937 & 278.4 & 11937 \\
    100-5-2 & 100   & 5     & 5     & 32    & 10    & 10171.0 & 10157 & 393.3 & 10185.1 & 10157 & 405.2 & 11282.4 & 11189 & 393.4 & 10578.0 & 10578 & 32.9  & 10578 \\
    100-5-2b & 100   & 5     & 5     & 15    & 10    & 7814.0 & 7814  & 221.5 & 8032.0 & 7833  & 165.4 & 8833.7 & 8657  & 221.6 & 8426.1 & 8307  & 429.4 & 8307 \\
    100-5-3 & 100   & 5     & 5     & 30    & 10    & 10451.0 & 10451 & 77.5  & 10458.1 & 10451 & 213.9 & 10876.7 & 10863 & 77.6  & 10651.4 & 10651 & 267.7 & 10651 \\
    100-5-3b & 100   & 5     & 5     & 16    & 10    & 8283.0 & 8283  & 119.5 & 8287.1 & 8283  & 167.7 & 9341.6 & 9332  & 119.6 & 9068.0 & 9063  & 244.8 & 9018 \\
    100-10-1 & 100   & 10    & 5     & 35    & 11    & 11297.0 & 11247 & 424.2 & 11268.2 & 11247 & 129.5 & 11963.8 & 11942 & 424.3 & 11451.8 & 11409 & 435.4 & 11409 \\
    100-10-1b & 100   & 10    & 5     & 18    & 11    & 9243.6 & 9151  & 549.3 & 9257.2 & 9242  & 163.3 & 10374.0 & 10346 & 549.4 & 10194.3 & 10168 & 239.5 & 10168 \\
    100-10-2 & 100   & 10    & 5     & 33    & 11    & 10140.8 & 10100 & 336.5 & 10158.0 & 10127 & 336.6 & 10879.0 & 10829 & 336.6 & 10561.5 & 10525 & 395.9 & 10515 \\
    100-10-2b & 100   & 10    & 5     & 18    & 11    & 7825.6 & 7781  & 343.4 & 7965.2 & 7956  & 257.5 & 9558.3 & 9463  & 343.5 & 8800.7 & 8752  & 284.0 & 8621 \\
    100-10-3 & 100   & 10    & 5     & 32    & 11    & 10546.9 & 10503 & 318.4 & 10530.1 & 10490 & 180.6 & 11256.4 & 11164 & 318.5 & 10743.6 & 10730 & 276.3 & 10730 \\
    100-10-3b & 100   & 10    & 5     & 17    & 11    & 8646.2 & 8554  & 448.7 & 8683.3 & 8628  & 269.2 & 9503.2 & 9414  & 448.8 & 9209.6 & 9144  & 245.2 & 9144 \\
    200-10-1 & 200   & 10    & 5     & 62    & 20    & 15918.6 & 15615 & 796.0 & 15544.6 & 15453 & 362.5 & 16696.7 & 16426 & 796.1 & 16354.8 & 16016 & 417.1 & 16016 \\
    200-10-1b & 200   & 10    & 5     & 30    & 20    & 12310.7 & 11871 & 730.7 & 12076.1 & 11908 & 338.5 & 13510.3 & 13229 & 730.8 & 12975.4 & 12771 & 571.8 & 12768 \\
    200-10-2 & 200   & 10    & 5     & 63    & 20    & 13986.3 & 13669 & 635.4 & 13696.4 & 13577 & 263.7 & 14087.1 & 13971 & 635.5 & 14132.9 & 13860 & 329.2 & 13860 \\
    200-10-2b & 200   & 10    & 5     & 30    & 20    & 10089.7 & 10025 & 504.2 & 10257.7 & 10004 & 363.0 & 11099.6 & 10837 & 504.3 & 10833.0 & 10515 & 355.4 & 10495 \\
    200-10-3 & 200   & 10    & 5     & 63    & 20    & 18102.1 & 18000 & 580.8 & 17990.0 & 17925 & 622.6 & 18251.8 & 18167 & 580.9 & 18144.4 & 18094 & 443.9 & 18073 \\
    200-10-3b & 200   & 10    & 5     & 30    & 20    & 12088.4 & 12021 & 500.4 & 12001.3 & 11955 & 380.0 & 12711.4 & 12586 & 500.5 & 12515.3 & 12445 & 468.0 & 12441 \\
    \addlinespace[5pt]
		Average &       &       &       &       &       & 11318.5 & 11220.2 & 421.4 & 11292.2 & 11221.8 & 281.4 & 12206.6 & 12083.5 & 421.5 & 11829.8 & 11729.6 & 339.9 & 11716.4 \\
    \bottomrule
    \end{tabular}%
\end{table}%
\end{landscape}

\noindent
\textbf{Impact of the main LNS components.}
We conducted additional experiments to measure the contribution of each operator of the LNS-E2E. Starting from the current algorithm (baseline configuration), we deactivated one separate destroy operator listed in Section \ref{sec:destroy}, in turn, and measured the solution quality of resulting algorithm. In the specific case of the configuration ``No open'', all candidate satellites are made available again (re-opened) at each restart, instead of using this component as a destroy operator. These experiments were conducted on the 18 large instances of Set~5, using 10 independent runs and a time limit $T_\textsc{max} = 15$ minutes. The solution quality is reported as an average percentage gap from the baseline. Table~\ref{tab:components} summarizes the results.

\begin{table}[htbp]
\renewcommand{\arraystretch}{1.2}
\small
\tabcolsep=10pt
\centering
  \caption{Sensitivity analysis on the contribution of each operator.\label{tab:components}}
	\begin{tabular}{rlllll}
    \toprule
    Baseline  & A) No related & B) No route & C) No close & D) No open & E) No single  \\
    \midrule
    11829.8 & 2.70\% & 2.44\%& 1.52\% & 3.10\% & 1.61\%  \\
    \bottomrule
    \end{tabular}
\end{table}%

From these results, we first observe that the ``open all satellites'' operator (D) is essential for the performance of the method, as it allows to control the frequency of the exploration of different satellite configurations. Without the explicit use of a dedicated operator to re-open satellites, the solutions are 3.10\% worse on average. The operator ``closes satellite'' (C) has a smaller but still very significant impact on the overall solution quality, with a deviation of 1.52\% from the baseline when deactivated. Therefore, forcing the elimination of some satellite choices at different phases of the method is essential to evaluate structurally different solutions which would not be attained otherwise by the cheapest insertion repair heuristic.

\emph{No related} measures the deterioration due to the deactivation of the related nodes destruction operator (A), which destroys specific areas around a seed customer. Analogously, \emph{no route} measures the performance deterioration when the operator targeting random routes (B) is deactivated, and column \emph{no single} shows the impact of not using the destruction operator which removes single-customer routes (E). All these operators appear to contribute significantly to the performance of the method, and the deactivation of any of these elements leads to an overall drop of method performance.

We finally tested a version of the method without a restart process after each $i_{max}$ iterations without improvement. In this configuration, the loop of Algorithm 1, Line 4--8 is executed until reaching a maximum time of $T_\textsc{max}$. We observed a deterioration of solution quality of $1.90\%$ with respect to the baseline, demonstrating again the importance of diversification components, in this case the restarts mechanism, for the E2E-VRP.

\subsection{Sensitivity analysis -- Density of charging stations and battery capacity}
\label{section:analysis3}

Our second set of experiments focuses on the impact of two defining features of battery-powered distribution networks: the density of the charging stations, and the range of the vehicles.
To that extent, we created two additional sets of instances with $\numc=50$ customers and $\nums=4$ satellites each, approximating as closely as possible real delivery conditions in a metropolitan area while pertaining to the \acp{E2E-VRP} class. Set~7 contains 10 groups of 20 instances with a number of charging stations $\numr \in \{2,3,5,10,15,20,25,30,40,50\}$ (in addition to the satellites) and a battery capacity $L=1000$, whereas Set~8 contains 10 groups of 20 instances with $\numr=20$ charging stations and a battery capacity $L \in \{800,900,1000,\dots,1700\}$. When varying the number of charging stations or the battery capacity, all other instance characteristics (satellite locations, customers locations and demands as well as the existing charging stations locations) are kept identical.

In each instance, 40 customers have been located randomly (with uniform probability) in an ellipse $X_1$ centered in (1000,500), with x-axis of dimension 800 and y-axis of dimension 400, and 10 additional customers have been located randomly in an ellipse $X_2$ with the same center, an x-axis of dimension 1000 and y-axis of dimension 500. The locations of the satellites are picked randomly in the area formed by $X_2 - X_1$, and the depot is fixed in position (300,0).
Moreover, 80\% of the charging stations are randomly located in $X_1$, and $20\%$ in $X_2$. The demand quantity of each customer is randomly selected in [1,25]. Six \ft-level vehicles with capacity $Q_1 = 250$ are available, and ten \sd-level electric vehicle with capacity $Q_2 = 125$ are available at each satellite.

Considering that one distance unit in each instance corresponds to 0.1km on a map, the area considered for the location of customers and charging stations covers 15708 km$^2$, a size similar in magnitude with the metropolitan area of Paris. We set a baseline of $L = 1000$ for the battery capacity, equivalent to a range of 100km. This value matches the specifications of the Renault Kangoo Zoe and Nissan Leaf 2015/2016 minus a safety range of 30km.
Varying this parameter allows to evaluate the impact of the battery capacity.

\begin{figure}[htbp]
\centering
\includegraphics[width = 0.65\textwidth]{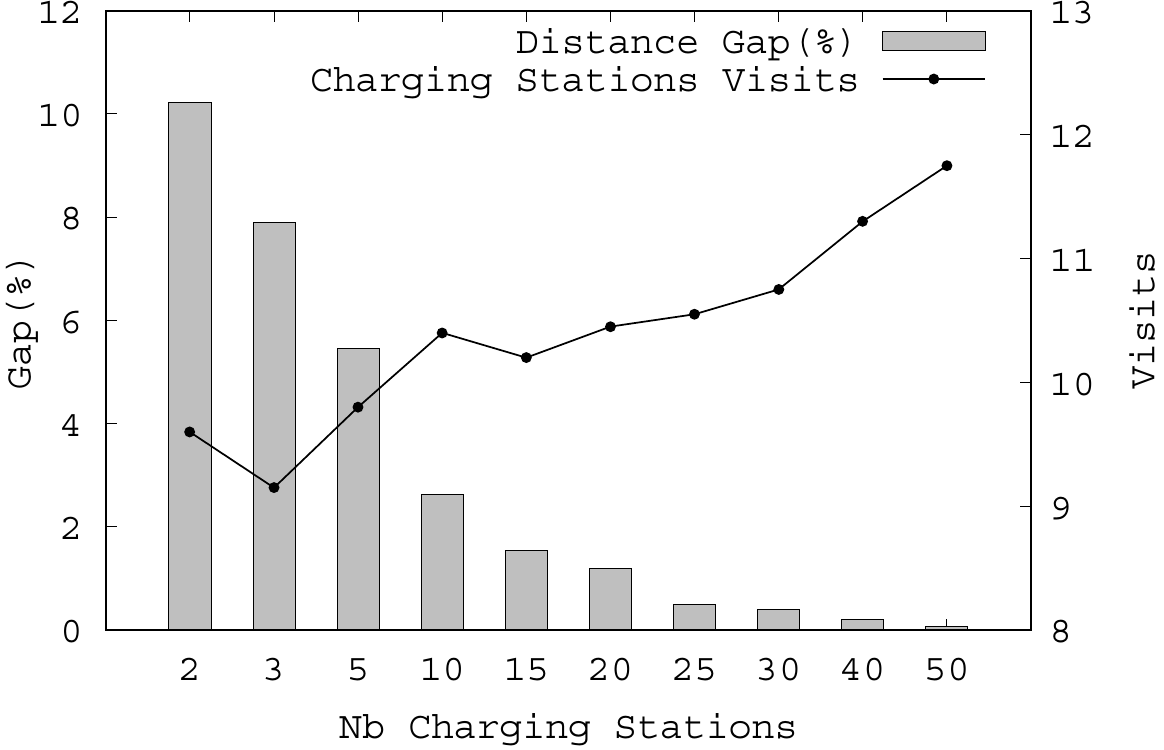}
\caption{Impact of the number of available charging stations on the detour costs due to recharging and the number of visits to stations.}
\label{Sensi-Analysis}
\end{figure}

\begin{figure}[htbp]
\centering
\includegraphics[width = 0.7\textwidth]{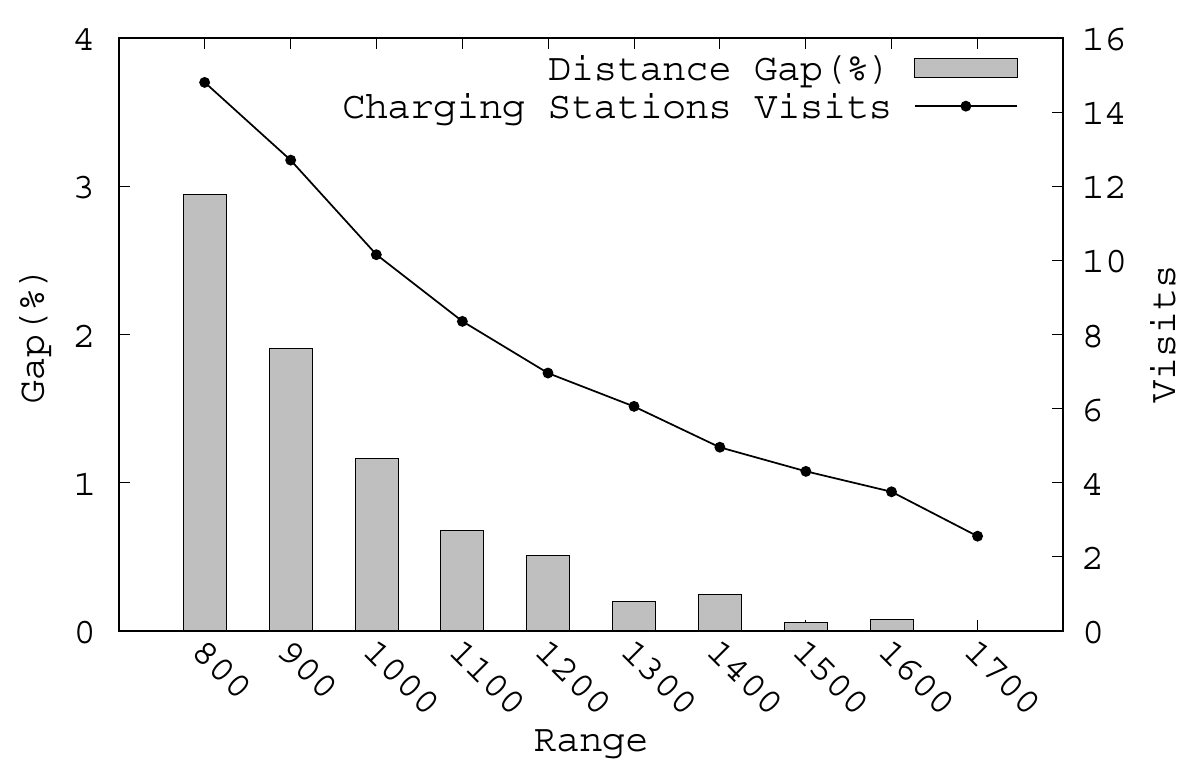}
\caption{Impact of the vehicle range (i.e., battery capacity) on the detour costs due to recharging and the number of visits to stations.}
\label{Sensi-Analysis2}
\end{figure}

Figure \ref{Sensi-Analysis} depicts the evolution of the operational costs as a function of the number of charging stations $\numr$, and Figure \ref{Sensi-Analysis2} shows the impact of the battery capacity $L$. The results are expressed as percentage gaps between the cost of the LNS-E2E solution with and without battery restrictions (i.e. percentage detour miles due to recharging), and averaged over all~20 instances of each class. The average number of visits to charging stations in the solutions is also represented on each graph.\vspace*{-0.2cm}

These experiments highlight the significant impact of the charging stations density and vehicles batteries capacities in the instances under study. As the number of charging stations grows, the detour costs due to recharging stations visits rapidly decreases: e.g., from 5.45\% in average when $\numr=5$ to 1.53\% when $\numr=15$. Conducting a power-law regression of the form $f(\numr) = \alpha / \numr^\beta$ (least-squares regression of an affine function on the log-log graph), the extra detours due to recharging diminish proportionally to $1/\rho^{1.24}$. In these conditions, doubling the number of charging stations allows to reduce extra recharging costs by approximately 58\%.

Interestingly, the \emph{number of visits} to charging stations slightly increases with $\numr$: from 9.5 in average when $\numr=2$ to 11.75 when $\numr=50$. Indeed, when the recharging station network is sparse, most detour options to recharging stations involve significant extra costs, and the vehicle routing algorithm tends to reduce to a strict minimum the number of such detours. In contrast, in the presence of a dense recharging stations network, the solutions converge more closely towards the 2E-VRP cost (disregarding battery constraints) as there are always multiple options of charging stations on the way.

The vehicles' range (i.e., battery capacity), has an even larger impact (see Figure \ref{Sensi-Analysis2}). For most of the considered instances, a range below $700$ distance units ($=$ 70km) would lead to an infeasible problem, as it becomes impossible to travel between some pairs of customers and find adequate recharging locations en-route. Therefore, adequate battery technology is a key factor for the viability of battery-powered delivery networks. The extra costs due to recharging and number of visits to recharging stations tend to be high when considering vehicles' ranges close to the feasible limit ($L=800$). These values then rapidly decrease to become close to zero when the range $L$ exceeds $1500$ (i.e., $150$km), a value which may be soon attained by lightweight electric trucks.
Once this regime is attained, the battery capacity becomes sufficient to do most tours without en-route recharging visits.
Still, manufacturing the current best-performing batteries on a global scale requires a large supply of minerals (e.g., nickel and cobalt) which are only accessible in limited quantities in the environment. As illustrated in Figure \ref{Sensi-Analysis}, the development of good fast-charging infrastructures is another strategic development path to obtain operational efficiency, which may turn out, in the long run, to be more sustainable and economical than a race towards heavier and more robust batteries.

\section{Conclusions}
\label{sec:conc}

In this paper, we have formulated the \ac{E2E-VRP}, an extension of the \ac{2E-VRP} involving electric vehicles for second-echelon deliveries, battery capacity constraints, and possible visits to charging stations, and used it as a prototypical problem for the study of multi-echelon battery-powered supply chains. We introduced an efficient exact algorithm, based on the enumeration of candidate solutions for the first echelon and on bounding functions and route enumeration for the second echelon, along with a problem-tailored large neighborhood search metaheuristic (LNS-E2E). A comparison of the solutions found by the LNS-E2E with lower bounds and optimal solutions produced by the mathematical programming algorithm demonstrates the excellent performance of both algorithms. In particular, all known optimal solutions for small instances were retrieved by the LNS-E2E, and an average optimality gap of 3.57\% between the best known upper and lower bounds was obtained on medium-scale instances. The metaheuristic was also evaluated on the classical \ac{2E-VRP} (without electric vehicles), producing new state-of-the-art solutions on large-scale instances with 200 customers. Finally, thanks to the use of efficient heuristic move filters in the local search and labeling algorithms, the computational effort of LNS-E2E remains comparable with that of previous metaheuristics for the classical \ac{2E-VRP}.

Beyond the usefulness of these optimization algorithms for the operational planning of electric fleets, this paper brought new managerial insights related to the incorporation of electric vehicles into two-echelon delivery networks and to the recharging-stations infrastructure required for an efficient supply chain. For this additional study, we created 400 additional test instances simulating typical requests patterns and delivery infrastructures in a metropolitan area with varying density of charging stations vehicles' battery capacities. We observed that the detour miles due to recharging decrease in $\mathcal{O}(1/\rho^{x})$ with $x \approx 5/4$ as a function of the number of charging stations. Moreover, the range of the electric vehicles has an even bigger impact: an increase of battery capacity to a range of 150km helps performing the majority of suburban delivery tours without need for en-route recharging, but a battery capacity below 80km render electric deliveries unviable in our setting. Between these two extremes, the extra costs due to recharging quickly decrease as a function of the battery capacity.

The future research perspectives are multiple.
Firstly, we recommend to pursue the study of exact methods and metaheuristics for multi-echelon electric \acp{VRP}. These optimization problems involve a large number of decision classes, related to satellite choices, recharging stations choices, and vehicle routing at two levels, posing a formidable challenge for exact and heuristic algorithms alike.
With the rapid development of battery-powered vehicles and green supply chains, efficient algorithms for large scale problems are critically needed, but the current methods still need to be improved in terms of accuracy, scalability, and generality, e.g., considering possible extensions to multi-echelon electric delivery schemes arising in city logistics \citep{Cattaruzza2017a}, other vehicle routing attributes \citep{Vidal2013} and stochastic settings.

Secondly, our sensitivity analyses on electric vehicles characteristics and other strategic decisions (number and placement of charging stations) could be extended further. One limitation of the current study relates to the placement of the charging stations, which is randomized in a fixed area. However, during urban planning, recharging stations are placed in strategical locations to meet the needs of the population, or based on competitive location approaches. Solving this strategic location optimization problem may be necessary for a more accurate sensitivity analysis.
Yet, it is an intricate problem, which can be viewed as a variant of location routing problems \citep{Schiffer2017,Schiffer2017a}, or modeled as a bilevel optimization problem and congestion game \citep{Xiong2015}. To this date, the optimization of charging stations locations has never been considered in the context of a multi-echelon delivery network, forming a promising research avenue.

\section*{Acknowledgements}
The authors would like to thank the anonymous reviewers and associate editor for their helpful suggestions.
This study was partially funded by the Austrian Climate and Energy Fund within the ``Electric Mobility Flagship projects'' program under grant 834868 (project VECEPT) in Austria, and CNPq and FAPERJ in Brazil (grant numbers 308498/2015-1 and E-26/203.310/2016). This support is gratefully acknowledged.

\begin{acronym}
	\setlength{\itemsep}{-\parsep}																		
		\acro{2E-VRP}[2EVRP]{two-echelon vehicle routing problem}
		\acro{2E-LRP}[2ELRP]{two-echelon location routing problem}
		\acro{2E-LRPSD}[2ELRPSD]{two-echelon location routing problem with single depot}
		\acro{AIT}{Austrian Institute of Technology}
		\acro{ALNS}{adaptive large neighborhood search}
		\acro{BKS}{best known solution}
		\acro{cf}[cf.\ ]{confer}
		\acro{CMA-ES}{covariance matrix adaptation evolution strategy}
		\acro{Cust}[Cust.\ ]{Customer}
		\acro{CVRP}{capacitated vehicle routing problem}
		\acro{CVRPSD}{capacitated vehicle routing problem with split deliveries}
		\acro{distrib}[distrib.\ ]{distribution}
		\acro{DP}{dynamic programming}
		\acro{EFSMFTW}[EFSMFTW]{electric fleet size and mix problem with fixed costs and time windows}
		\acrodef{eg}[e.g.] {example given}
		\acrodef{etc}[etc.]{et cetera}
		\acro{EVRPTW}{electric vehicle routing problem with time windows}
		\acro{EVRPTWRS}{electric vehicle routing problem with time windows and recharging stations}
		\acrodef{ie}[i.e.]{id est}
		\acro{E2E-VRP}[E2EVRP]{electric two-echelon vehicle routing problem}
		\acro{E-VRPTW}{electric vehicle routing problem with time windows and recharging stations} 
		\acro{GVRP}{green vehicle routing problem}
		\acro{Inst}[Inst.]{Instance}
		\acro{LNS}{large neighborhood search}
		\acro{LRP}{location routing problem}
		\acro{MDC-VRP}[MDCVRP]{mul\-ti-de\-pot ca\-pa\-ci\-ta\-ted ve\-hi\-cle rou\-ting pro\-blem}
		\acro{MIP}{mixed integer program}
		\acro{NP}{non-deterministic polynomial-time}
		\acro{p}[p.]{page}
		\acro{Sat}[Sat.]{Satellite}
		\acro{SD-VRP}[SDVRP]{split delivery vehicle routing problem}
		\acro{Sol}[Sol.]{Solution}
		\acro{SPPRC}{shortest path problem with resource constraints}
		\acrodefplural{SPPRC}[SPPRCs]{shortest path problems with resource constraints}
		\acro{T}[T.]{Time}
		\acro{TS}{tabu search}
		\acro{TSP}{tra\-vel\-ling sales\-man prob\-lem}
		\acro{TTRP}{truck and trailer routing problem}
		\acro{VLNS}{very large neighborhood search}
		\acro{VNS}{variable neighborhood search}
		\acro{VRP}{vehicle routing problem}
\end{acronym}

\section*{Bibliography}

\bibliographystyle{abbrvnat}
\bibliography{library}

\end{document}